\documentclass[preprin,showpacs,preprintnumbers,eqsecnum,amsmath,amssymb,twoside]{revtex4}
\usepackage{graphics,epsfig}
\usepackage{graphicx}
\usepackage{dcolumn}
\usepackage{bm}

\begin{document}
\baselineskip=0.5 cm
\title{{\bf Quasinormal ringing of black holes in Einstein aether theory}}

\author{Chikun Ding }
\thanks{ Email: dingchikun@163.com; Chikun\_Ding@huhst.edu.cn }

  \affiliation{Department of Physics, Hunan University of Humanities, Science and Technology, Loudi, Hunan
417000, P. R. China\\
Key Laboratory of Low Dimensional
Quantum Structures and Quantum Control of Ministry of Education,
and Synergetic Innovation Center for Quantum Effects and Applications,
Hunan Normal University, Changsha, Hunan 410081, P. R. China}

\vspace*{0.2cm}
\begin{abstract}
\baselineskip=0.5 cm
\begin{center}
{\bf Abstract}
\end{center}

The gravitational consequence of local Lorentz violation (LV) should show itself in derivation of the characteristic quasinormal ringing of black hole mergers from their general relativity case.
In this paper, we study quasinormal modes (QNMs) of the scalar and electromagnetic field perturbations to Einstein aether black holes. We find that quasinormal ringing of the first kind aether black hole is similar to that of another Lorentz violation model---the QED-extension limit of standard model extension. These similarities between completely different backgrounds may imply that LV in gravity sector and LV in matter sector have some connections between themself: damping quasinormal ringing of black holes more rapidly and prolonging its oscillation period.  By compared to Schwarzschild black hole, both the first and the second kind aether black holes have larger damping rate and smaller real oscillation frequency of QNMs. And the differences are from 0.7 percent to 35 percents, those could be detected by new generation of gravitational antennas.

\end{abstract}

 \maketitle

\vspace*{0.2cm}

\section{Introduction}

After the first discovery of gravitational wave (GW) on September 14, 2015 (GW150914) \cite{abbott},  Laser Interferometer
Gravitational wave Observatory (LIGO) has detected GW for the third time, on January 4, 2017 (GW170104) \cite{driggers}. It provides a direct confirmation for the existence of a black hole and, confirms that black hole mergers are common in the universe, and will be observed in large numbers in the near future. The detections of GW give us opportunity and the ideal tool to stress test general relativity (GR) \cite{abbott2}. Some of them are used to test alternative theories of gravity where Lorentz invariance (LI) is broken which affects the dispersion relation for GW \cite{driggers}. For the first time, they used GW170104 to put upper limits on the magnitude of Lorentz violation tolerated by their data and found that the bounds are important.

Why consider Lorentz violation (LV)? Because that Lorentz invariance may not be an exact symmetry at all energies  \cite{mattingly}. Condensed matter physics, which has an analog of LI, suggests some scenarios of LV: a) LI is an approximate symmetry emerging at low energies and violated at ultrahigh energies \cite{chadha}; b) LI is fundamental but broken spontaneously \cite{klinkhamer}.
Any effective description must break down at a certain cutoff scale, which signs the emergence of new physical degrees of freedom beyond that scale. Examples of this include the hydrodynamics, Fermi's theory of beta decay \cite{bhattacharyya2} and quantization of GR \cite{shomer} at energies beyond the Planck energy. Lorentz invariance also leads to divergences in quantum field theory which can be cured with a short distance of cutoff that breaks it \cite{jacobson}.

 Thus, the study of LV is a valuable tool to probe the foundations of GR without preconceived notions of the numerical sensitivity \cite{shao}. These studies include LV in the neutrino sector \cite{dai}, the standard-model extension \cite{colladay}, LV in the non-gravity sector \cite{coleman}, and LV effect on the formation of atmospheric showers \cite{rubtsov}.
A more recent area for searching for LV is in the pure gravity sector, such as gravitational Cerenkov radiation \cite{kostelecky2015} and gravitational wave dispersion \cite{kostelecky2016}.
Einstein-aether theory can be considered as an effective description of Lorentz symmetry breaking in the gravity sector and has been extensively used in order to obtain quantitative constraints on Lorentz-violating gravity\cite{jacobson2}. On another side, violations of Lorentz symmetry have been used to construct modified-gravity theories that account
for dark-matter phenomenology without any actual dark mater \cite{bekenstein2004}.

Einstein-aether theory \cite{jacobson2}  is originated from the scalar-tensor theory \cite{gong}. In Einstein-aether theory, the background tensor fields break the Lorentz symmetry only down to a rotation subgroup by the existence of a preferred time direction at every point of spacetime, i.e., existing a preferred frame of reference established by aether vector $u^a$. The introduction of the aether vector allows for some novel effects, e.g., matter fields can travel faster than the speed of light \cite{jacobson3}, dubbed superluminal particle. It is the
 universal horizons that can trap excitations traveling at arbitrarily high velocities.
In 2012, two exact black hole solutions and some mechanics of universal horizons in Einstein aether theory were found by Berglund {\it et al} \cite{berglund2012}. In 2015, two exact charged black hole solutions and their Smarr formula on both universal and Killing horizons were found by Ding {\it et al} \cite{ding}. In 2016, two exact black hole solutions and their Smarr formula on universal horizons in 3-dimensional spacetime were found by Ding {\it et al} \cite{ding2}. Other studies on universal horizons can be found in \cite{ding2016,tian}.

In Ref. \cite{ding2016}, Ding {\it et al} studied Hawking radiation from the charged Einstein aether black hole and found that i) the universal horizon seems to be no role on the process of radiating luminal or subluminal particles; while ii) the Killing horizon seems to be no role on superluminal particle radiation. Since up to date, the particles with speed higher than vacuum light speed aren't yet found, we here consider only subluminal or luminal particles perturbation to these LV black holes. In 2007, Konoplya {\it et al} \cite{konoplya} studied the perturbations of the non-reduced Einstein aether black holes and found that both the real part and the absolute imaginary part of QNMs increase with the aether coefficient $c_1$.

Our goal here is to study on a perturbed black hole in Einstein aether theory. Perturbations of black holes in GR or alternative theories of gravity carry signatures of the effective potential around them and one could look for them. Once a black hole is perturbed, it responds to perturbations by emitting GWs \cite{konoplya2011} which are dominated by quasinormal ringing.  The GW signal can in general be divided into three stages: (i) a prompt response at early
times, which depends strongly on the initial conditions; (ii) an exponentially decaying ``ringdown" phase at intermediate times, where quasinormal modes
(QNMs) dominate the signal, which depends entirely on the final black hole's parameters and (iii) a late-time tail \cite{berti}. QNMs can be used in the analysis of a gravitational wave signal to provide a wealth of information: the masses and radii of the perturbed objects \cite{Nollert}. Recent QNMs study show that at the current precision of GW detections, there remains some possibility for alternative theories of gravity \cite{konoplya2016}.

So by studying these LV black holes' QNMs, we can obtain some signal of LV from future GW events. By using QED-extension limit of standard model extension ( SME, see Appendix for more detail), Chen {\it et al} \cite{chen2006} has studied the influence of LV on Dirac field perturbation to Schwarzschild black hole, and one will find that its properties have some similarities to our result. The plan of rest of our paper is organized as follows. In Sec. II we review briefly the Einstein aether black holes and the third order WKB method (A recent study on semianalytic technique appears in \cite{matyjasek}). In Sec. III we adopt to
the third order WKB method and obtain the perturbation frequencies of the first kind Einstein aether black holes. In Sec. IV, we discuss the QNMs for the second kind Einstein aether black hole. In Sec. V we present a summary. Appendix is for introducing SME and the accuracy of WKB method.

\section{Einstein aether black holes and WKB method}
The general action for the Einstein-aether theory can be constructed by assuming that: (1) it is general covariant; and (2)  it  is a functional of only the spacetime metric $g_{ab}$ and a unit timelike vector $u^a$, and involves  no more than two derivatives of them, so that the resulting field equations are second-order differential equations of  $g_{ab}$ and   $u^a$. Then,  the  Einstein aether theory to be studied   in this paper is
 described by the  action,
\begin{eqnarray}
\mathcal{S}=
\int d^4x\sqrt{-g}\Big[\frac{1}{16\pi G_{\ae}}(\mathcal{R}+\mathcal{L}_{\ae})\Big], \label{action}
\end{eqnarray}
where $G_{\ae}$ is the aether gravitational constant, $\mathcal{L}_{\ae}$ is the aether Lagrangian  \begin{eqnarray}
-\mathcal{L}_{\ae}=Z^{ab}_{~~cd}(\nabla_au^c)(\nabla_bu^d)-\lambda(u^2+1)
\end{eqnarray}
with
\begin{eqnarray}
Z^{ab}_{~~cd}=c_1g^{ab}g_{cd}+c_2\delta^a_{~c}\delta^b_{~d}
+c_3\delta^a_{~d}\delta^b_{~c}-c_4u^au^bg_{cd}\,,
\end{eqnarray}
where $c_i (i = 1, 2, 3, 4)$ are coupling constants of the theory.
The aether Lagrangian is therefore the sum of all possible terms for the aether field $u^a$ up to mass dimension two, and the constraint term $\lambda(u^2 + 1)$ with the Lagrange multiplier $\lambda$ implementing the normalization condition $u^2=-1$.
There are a number of theoretical and observational bounds on the coupling constants $c_i$ \cite{jacobson2,yagi,jacobson5}. Here,  we impose the following
constraints\footnotemark\footnotetext{Note the slight difference between the constraints imposed here and the ones imposed in \cite{berglund2012}, as in this paper we also require that
vacuum Cerenkov radiation of gravitons is forbidden \cite{EMS}.},
\begin{equation}
\label{CDs}
0\leq c_{14}<2,\quad 2+c_{13}+3c_2>0,\quad 0\leq c_{13}<1,
\end{equation}
where $c_{14}\equiv c_1+c_4$, and so on.

The static, spherically symmetric metric for Einstein aether black hole spacetime can
be written in the form
\begin{eqnarray}
ds^2 = -f(r)\,dt^2 + \frac{dr^2}{f(r)} + r^2(d\theta^2
+\sin^2\theta d\phi^2)\,. \label{metric}
\end{eqnarray}
There are two kinds of exact solutions \cite{berglund2012,ding}. In the first case $c_{14}=0,\;c_{123}\neq0$ (termed the first kind aether black hole), the metric function is
\begin{eqnarray}
\label{sol1} f(r)=1-\frac{2M}{r}-I\Big(\frac{2M}{r}\Big)^4,\;\;I=\frac{27c_{13}}{256(1-c_{13})}.
\end{eqnarray}
If the coefficient $c_{13}=0$, then it reduces to Schwarzschild black hole. The quantity $M$ is the mass of the black hole spacetime\footnotemark\footnotetext{The total mass of the given spacetime is $MG_{\ae}=(1-c_{14}/2)r_0/2$. And the constant $G_{\ae}$ is related to Newton's gravitational constant $G_{N}$ by $G_{\ae}=(1-c_{14}/2)G_N$, which can be
obtained by using the weak field/slow-motion limit of the Einstein-aether theory \cite{carroll,eling,ding}. Therefore we can always set $r_0=2MG_N$ regardless of the coefficient $c_{14}$.}.
Its location of the Killing horizon is the largest root of $f(r)=0$, which is given by \cite{ding}
\begin{eqnarray}
 &&r_{KH}=M\left(\frac{1}{2}+L
 +\sqrt{N-P+\frac{1}{4L}}\right),\quad L=\sqrt{\frac{1}{4}+P},\quad
 \nonumber\\
 && P=\frac{2^{1/3}\cdot4I}{H}+\frac{H}{3\cdot2^{1/3}},\quad
 H=\left(27I+3\sqrt{3}I\sqrt{27-256I}\right)^{1/3}.
\end{eqnarray}

In the second case $c_{14}\neq0,\;c_{123}=0$ (termed the second kind aether black hole), the metric function is
\begin{eqnarray}
\label{sol2} f(r)=1-\frac{2M}{r}-J\Big(\frac{M}{r}\Big)^2,\;\;J=\frac{c_{13}-c_{14}/2}{1-c_{13}}.
\end{eqnarray}
Its Killing horizon locates at
\begin{eqnarray}
 &&r_{KH}=M\left(1+\sqrt{J+1}\right).
\end{eqnarray}
If the coefficient $c_{13}=c_{14}/2$, it also reduces to Schwarzschild black hole.

There is an universal horizon in these black hole spacetimes behind their Killing horizons even for aether coefficient $c_{13}=0$ or $c_{13}=c_{14}/2$ (see \cite{berglund2012,ding} for more detail). The universal horizon can trap particles with arbitrary high velocity, i.e., super-luminal particles. The killing horizons are invisible to these super-luminal particles. In another side, the universal horizon seems has no role on radiating luminal or sub-luminal particles during Hawking radiation \cite{ding2016}. For this reason, to luminal or sub-luminal particles perturbation, we here wouldn't consider the role of the universal horizon at present.

To scalar and electromagnetic fields perturbation, we shall neglect interaction of these fields with aether for simplicity and use general covariant wave equations.
Then, the wave equations for test scalar $\Phi$ and electromagnetic $A_\mu$ fields are
\begin{eqnarray}
&&\frac{1}{\sqrt{-g}}\partial_\mu(\sqrt{-g}g^{\mu\nu}\partial_\nu\Phi)=0,\nonumber\\
&&\frac{1}{\sqrt{-g}}\partial_\mu(\sqrt{-g}F^{\mu\nu})=0,
\end{eqnarray}
with $F_{\mu\nu}=\partial_\mu A_\nu-\partial_\nu A_\mu$.
They can be reduced to Schrodinger like equations:
\begin{eqnarray}\label{schrodinger}
\frac{d^2\Psi_i}{dr_*^2}+[\omega^2-V_i(r)]\Psi_i=0,\;\;dr_*=f(r)dr,
\end{eqnarray}
for scalar field $\Psi_s$ and electromagnetic one $\Psi_e$.
The effective potentials take the form as:
\begin{eqnarray}\label{potential}
V_i=f(r)\left[\frac{l(l+1)}{r^2}+\frac{\beta}{r}\frac{df(r)}{dr}\right],
\end{eqnarray}
where $\beta=1$ for the scalar field potential $V_s$, $\beta=0$ for the electromagnetic one $V_e$, respectively. The effective potentials $V_i$ depend on the value $r$, angular quantum number (multipole momentum) $l$ and the aether coefficient $c_{13}$.

\begin{figure}[ht]
\begin{center}
\includegraphics[width=5.0cm]{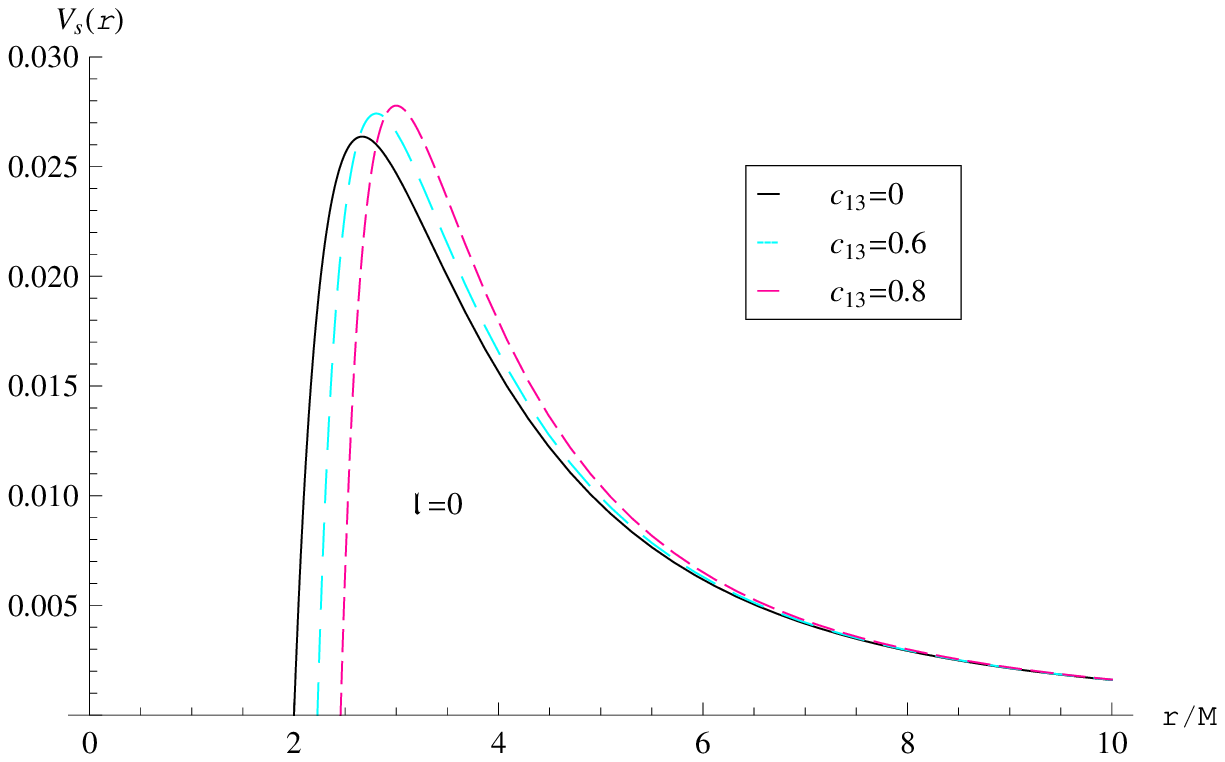}\;\;\includegraphics[width=5.0cm]{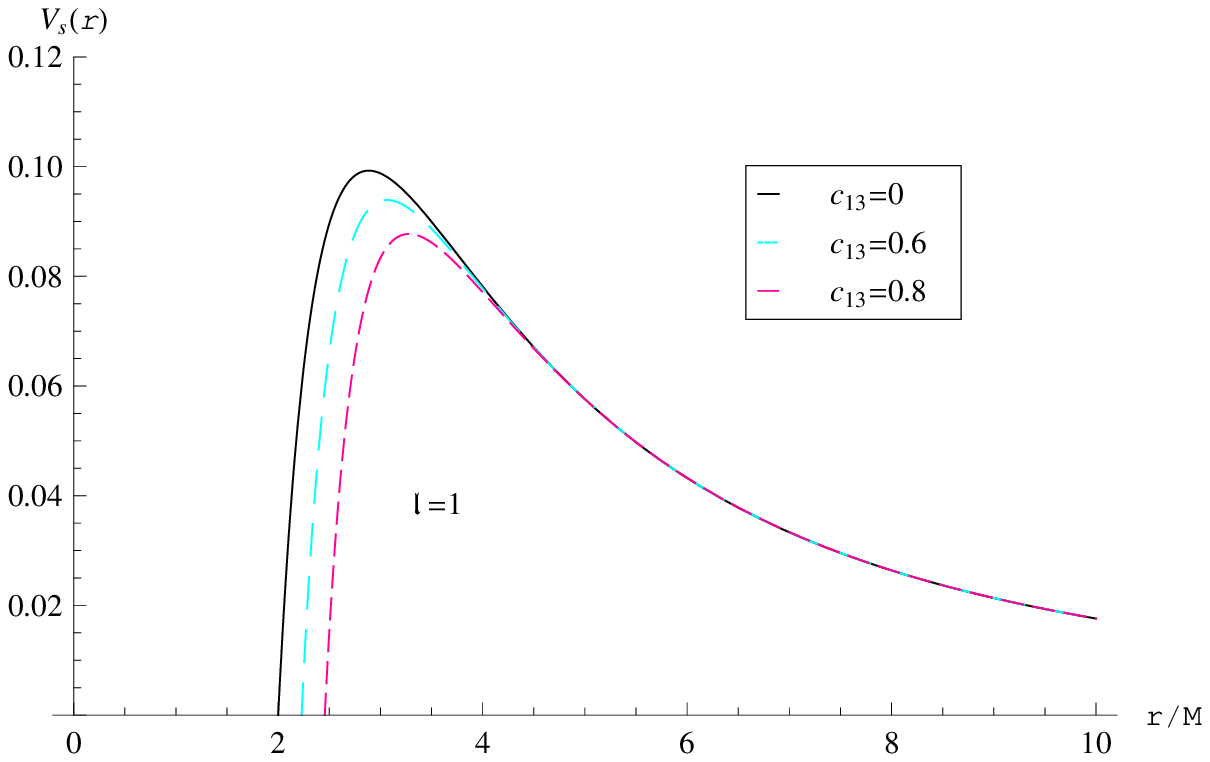}\;\;
\includegraphics[width=5.0cm]{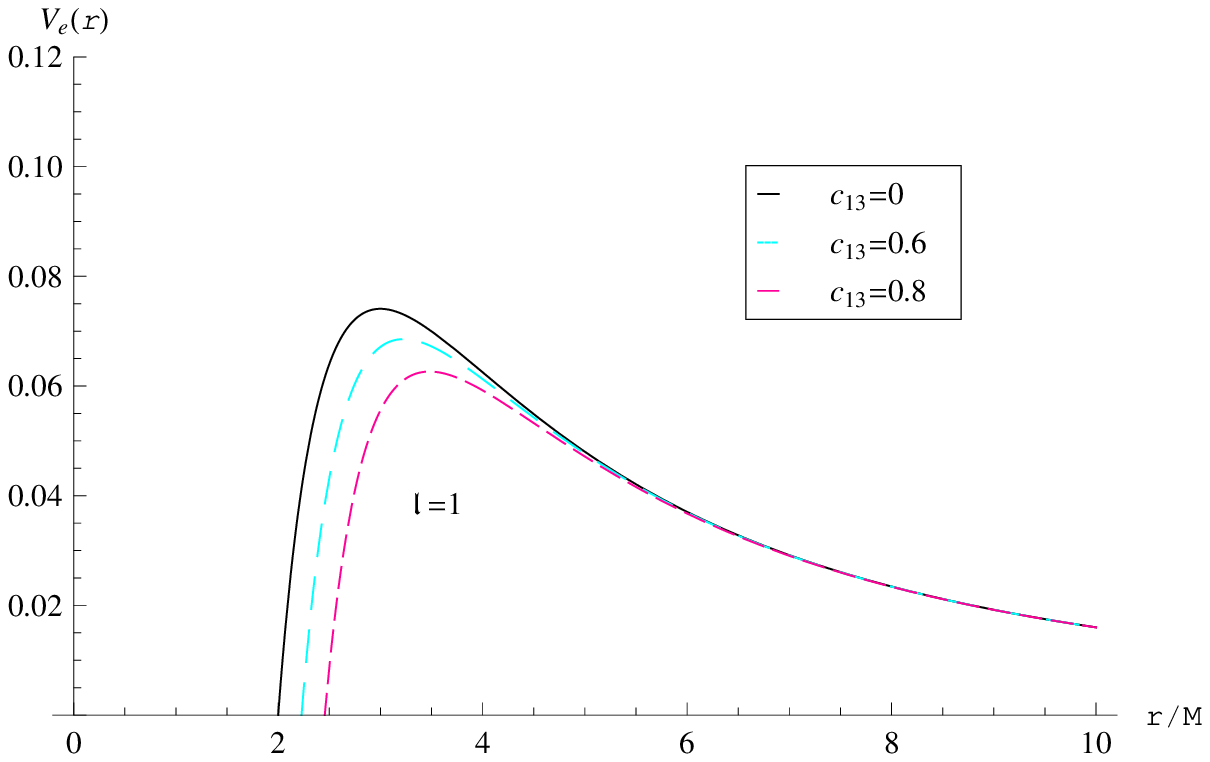}
\caption{The left both figures are the effective potential of scalar field perturbations $V_s$ near the first kind aether black hole $(M=1)$ with different coefficients $c_{13}$. The third figure is for the electromagnetic field perturbations $V_e$.}\label{fp1}
 \end{center}
 \end{figure}

From the potential formula (\ref{potential}), the effective potential for the first kind aether black hole is
\begin{eqnarray}
V_i=\big(1-\frac{2M}{r}\big)\left[\frac{l(l+1)}{r^2}+\frac{2M\beta}{r^3}\right]
+\frac{16M^4I}{r^6}\left[2\beta\big(2-\frac{5M}{r}-32I\frac{M^4}{r^4}\big)-l(l+1)\right],
\end{eqnarray}
where the the first two terms are Schwarzschild potential, the rests are the aether modified terms, shown in Fig. \ref{fp1}.
In FIG. \ref{fp1},  it is the effective potential of scalar and electromagnetic  field perturbations near the first kind aether black hole. Obviously, if $c_{13}=0$, the effective potentials $V_i$ can be reduced to those of the Schwarzschild black hole. For $l=0$, the peak value of the scalar potential barrier gets higher with $c_{13}$ increasing. On the contrary, for $l>0$ the peak value gets lower with $c_{13}$. This contrariness is similar to the case of the deformed Ho\v{r}ava-Lifshitz black hole \cite{chen2010} where the peak gets lower for $l=0$ and higher for $l>0$ with the parameter $\alpha$ increase. In Einstein-Maxwell theory, i.e., Reissner-Norstr\"{o}m black hole, the electric charge $Q$ increases the peak for all $l$. In the Einstein-Born-Infeld theory, the Born-Infeld scale parameter $b$ decreases the peak for all $l$. These properties of the potential will imply that the quasinormal modes posses some different behavior from those black holes.

\begin{figure}[ht]
\begin{center}
\includegraphics[width=5.0cm]{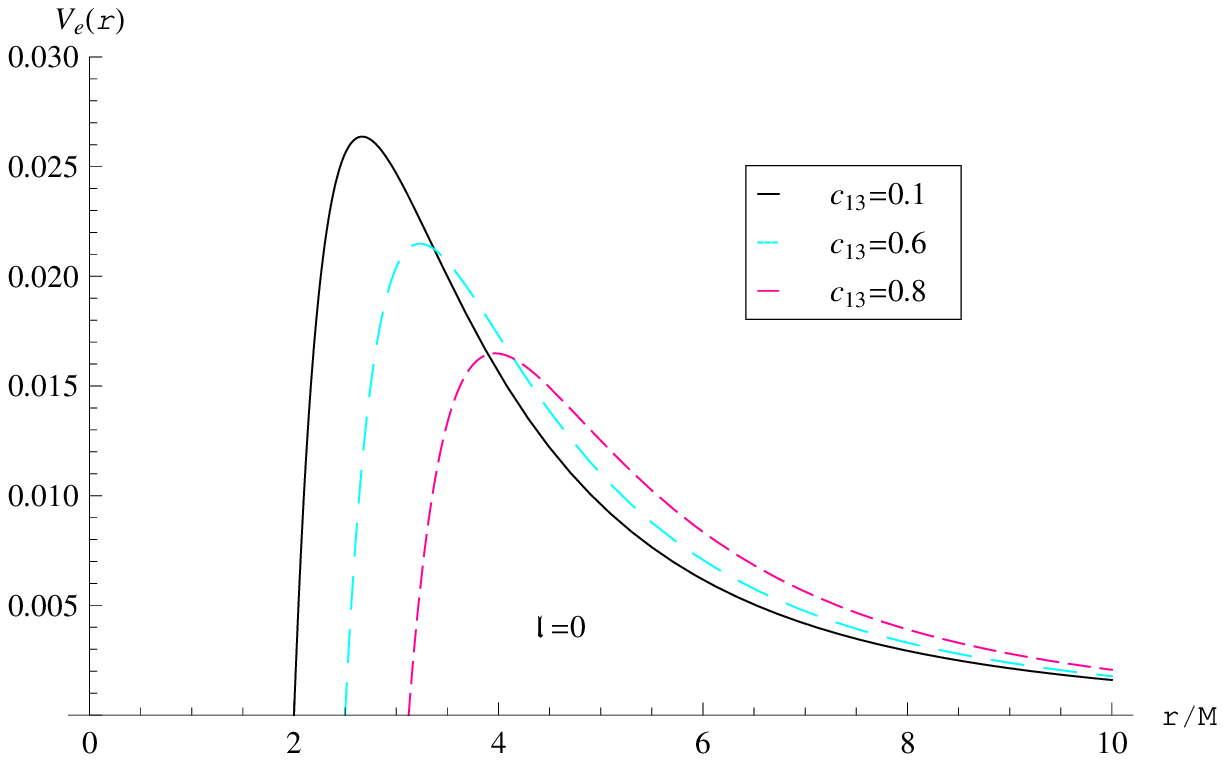}\;\;\includegraphics[width=5.0cm]{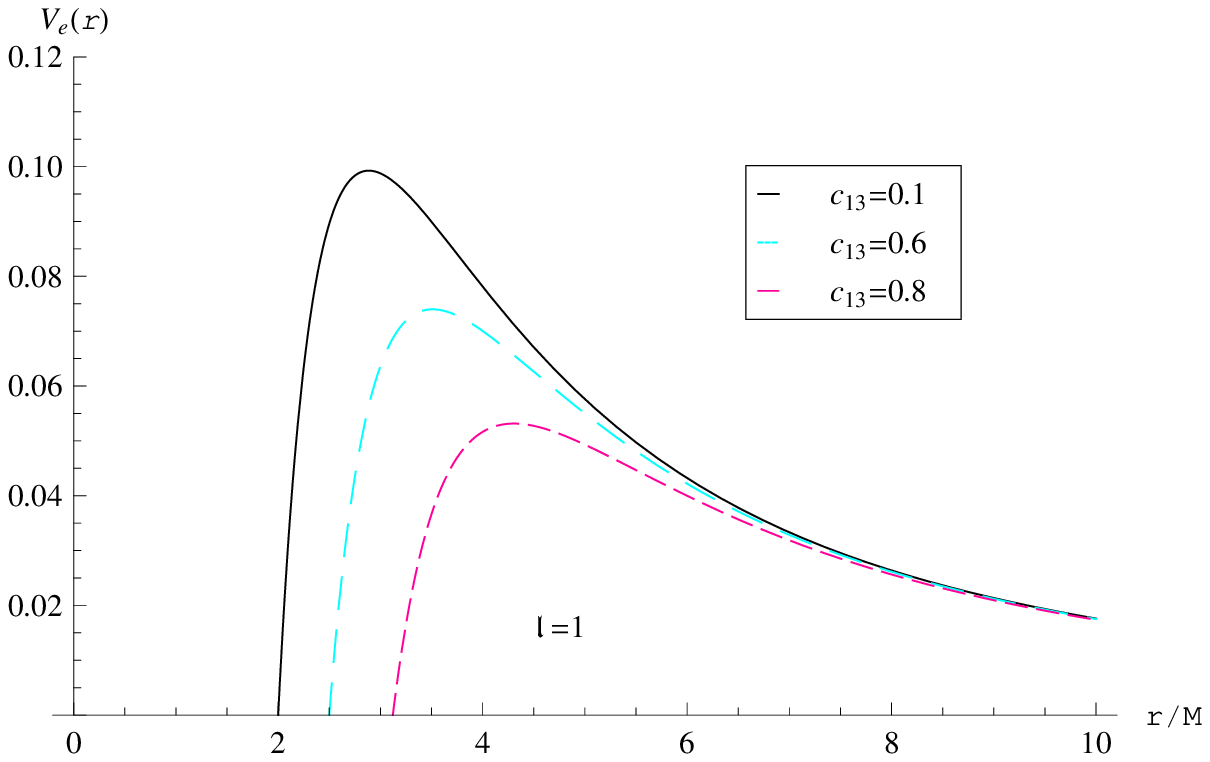}\;\;
\includegraphics[width=5.0cm]{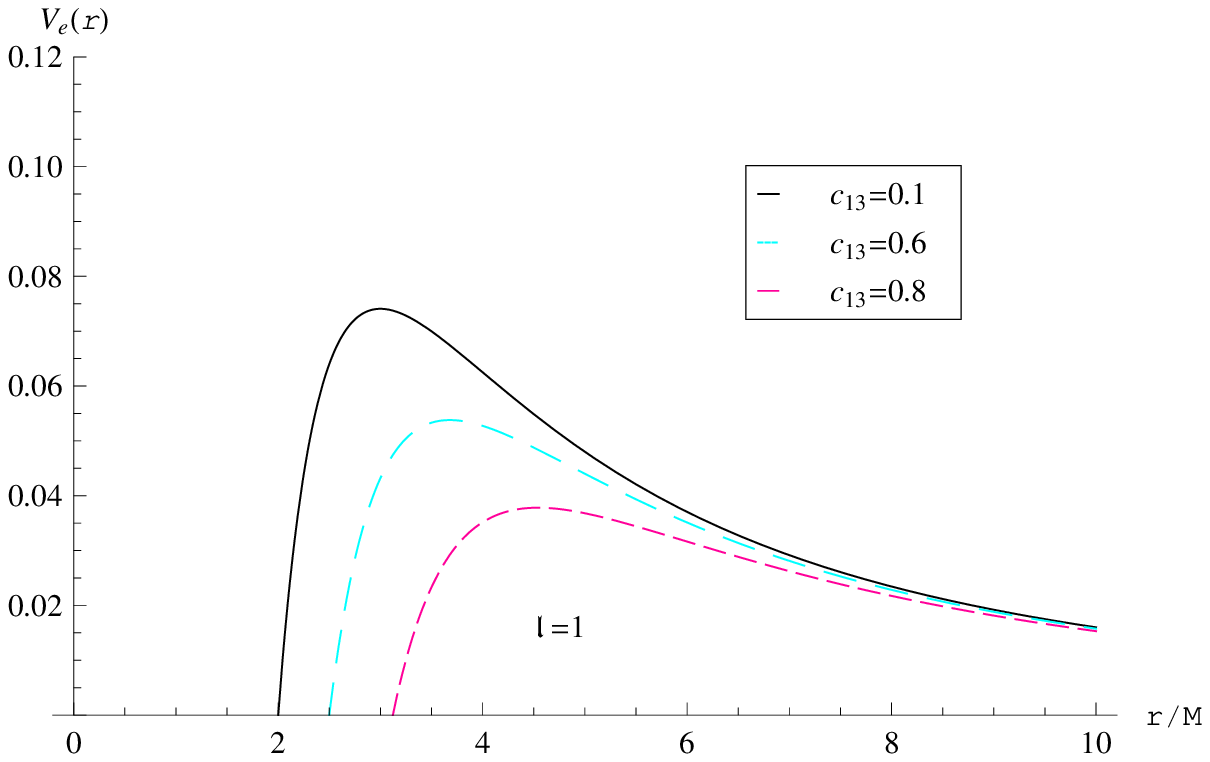}
\caption{The left both figures are the effective potential of scalar field perturbations $V_s$ near the second kind aether black hole $(M=1)$ with different coefficients $c_{13}$ and fixed coefficient $c_{14}=0.2$. The third figure is for the electromagnetic field perturbations $V_e$.}\label{fp12}
 \end{center}
 \end{figure}

From the potential formula (\ref{potential}), the effective potential for the second kind aether black hole is
\begin{eqnarray}
V_i=\big(1-\frac{2M}{r}\big)\left[\frac{l(l+1)}{r^2}+\frac{2M\beta}{r^3}\right]
+\frac{M^2J}{r^4}\left[2\beta\big(1-\frac{3M}{r}-J\frac{M^2}{r^2}\big)-l(l+1)\right],
\end{eqnarray}
where the the first two terms are Schwarzschild potential, the rests are the aether modified terms, shown in Fig. \ref{fp12}.
In FIG. \ref{fp12},  it is the effective potential of the scalar and electromagnetic field perturbation near the second kind aether black hole. It is easy to see that for all $l$, the peak value of the potential barrier gets lower with $c_{13}$ increasing just like the Born-Infeld scale parameter $b$ in the Einstein-Born-Infeld theory.

The Schr\"{o}dinger-like wave equation (\ref{schrodinger}) with the effective potential
(\ref{potential}) containing the lapse function $f(r)$ related to the Einstein aether black holes is not solvable analytically. Since then we now use the third-order WKB approximation method to evaluate the quasinormal modes of massless scalar and electromagnetic field perturbation to the first and second kind aether black holes. This semianalytic method has been proved to be accurate up to around one percent for the real and the imaginary parts of the quasinormal frequencies for low-lying modes with $n<l$.  Due to its considerable accuracy for lower lying modes, this method has been used extensively in evaluating quasinormal frequencies of various black holes. In this approximation, the formula for the complex quasinormal frequencies in this approximation is given by \cite{schutz,iyer1,iyer2}
 \begin{eqnarray}
\omega^2=\left[V_0+\sqrt{-2V_0''}\Lambda\right]-i\Big(n+\frac{1}{2}\Big)\sqrt{-2V_0''}(1+\Omega),
\end{eqnarray}
where
 \begin{eqnarray}
&&\Lambda=\frac{1}{\sqrt{-2V_0''}}\left[\frac{1}{8}\Big(\frac{V_0^{(4)}}{V_0''}\Big)
\Big(\frac{1}{4}+\alpha^2\Big)-\frac{1}{288}\Big(\frac{V_0^{(3)}}{V_0''}\Big)^2(7+60\alpha^2)\right],
\nonumber\\
&&\Omega=\frac{1}{-2V_0''}\Big[\frac{5}{6912}\Big(\frac{V_0^{(3)}}{V_0''}\Big)^4
\Big(77+188\alpha^2\Big)-\frac{1}{384}\Big(\frac{V_0'''^2V_0^{(4)}}{V_0''^3}\Big)(51+100\alpha^2)
\nonumber\\
&&+\frac{1}{2304}\Big(\frac{V_0^{(4)}}{V_0''}\Big)^2(67+68\alpha^2)
\frac{1}{288}\Big(\frac{V_0'''V_0^{(5)}}{V_0''^2}\Big)(19+28\alpha^2)\nonumber\\
&&-\frac{1}{288}\Big(\frac{V_0^{(6)}}{V_0''}\Big)(5+4\alpha^2)\Big],
\end{eqnarray}
and
 \begin{eqnarray}
\alpha=n+\frac{1}{2},\;\;V_0^{(m)}=\frac{d^mV_i}{dr_*^m}\Big|_{r_*(r_p)},
\end{eqnarray}
$n$ is overtone number and $r_p$ is the turning point value of polar coordinate $r$ at which  the effective potential reaches its maximum (\ref{potential}). Substituting the effective potential $V_i$ (\ref{potential}) into the formula above, we can obtain the quasinormal frequencies for the scalar and electromagnetic field perturbations to Einstein aether black holes. In the next sections, we obtain the quasinormal modes for the both kinds of Einstein aether black holes and analyze their properties.

\section{Quasinormal modes for the first kind aether black hole}

\begin{figure}[ht]
\begin{center}
 \includegraphics[width=7cm]{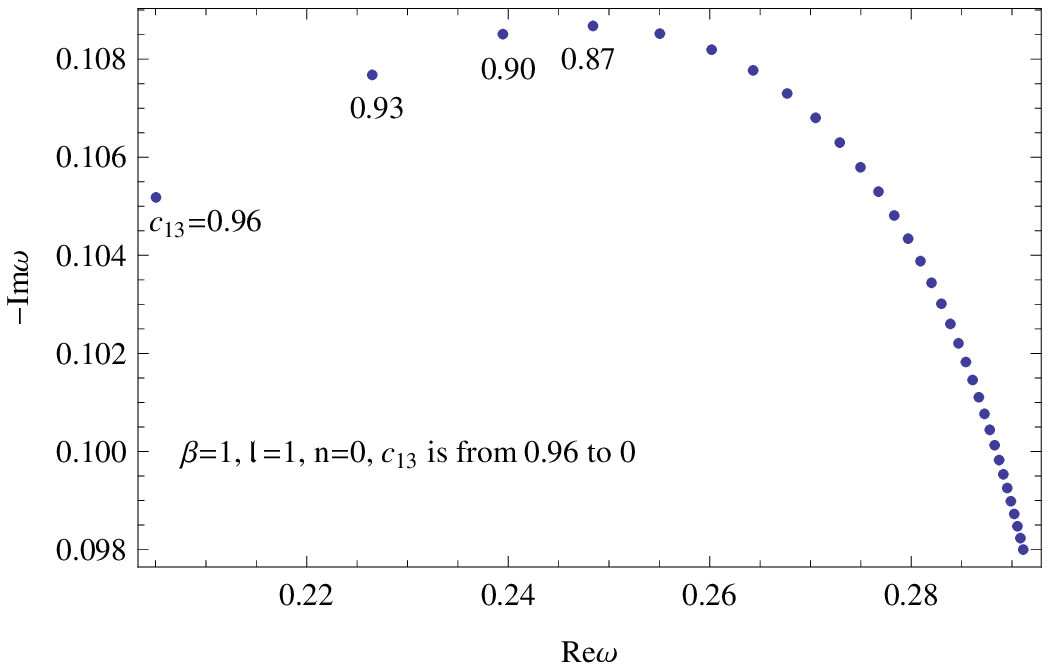}\;\;\;\;\includegraphics[width=7cm]{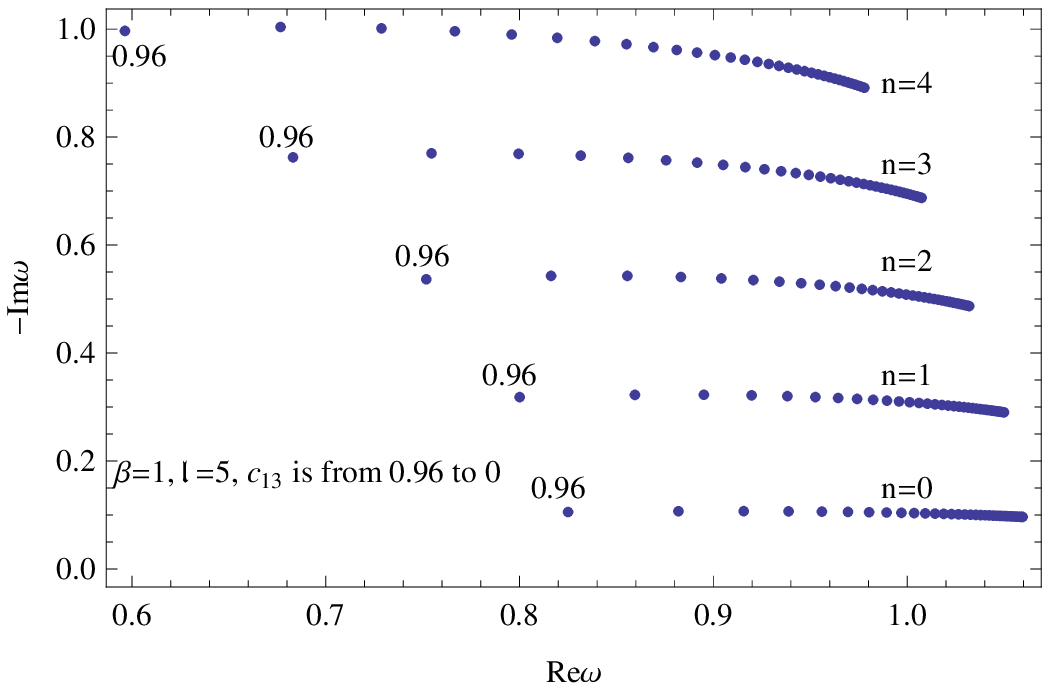}
\caption{The relationship between the real and imaginary parts of quasinormal frequencies of the scalar field in the background of the first kind aether black hole with the decreasing of $c_{13}$.}\label{fp2}
 \end{center}
 \end{figure}

\begin{figure}[ht]
\begin{center}
 \includegraphics[width=7cm]{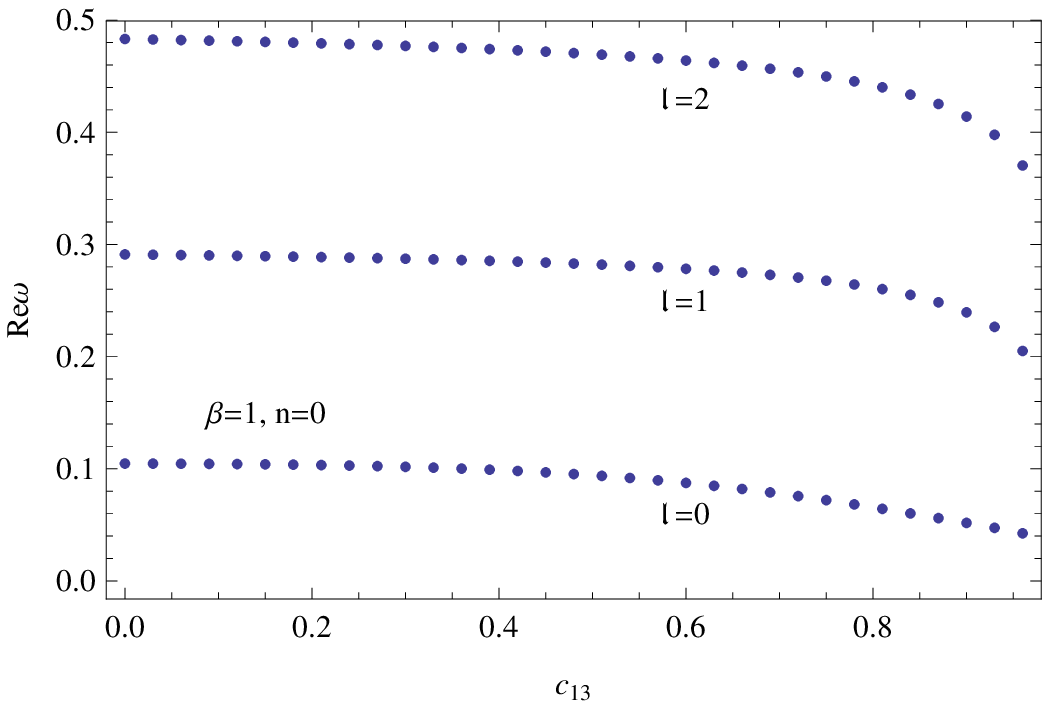}\;\;\;\;\includegraphics[width=7cm]{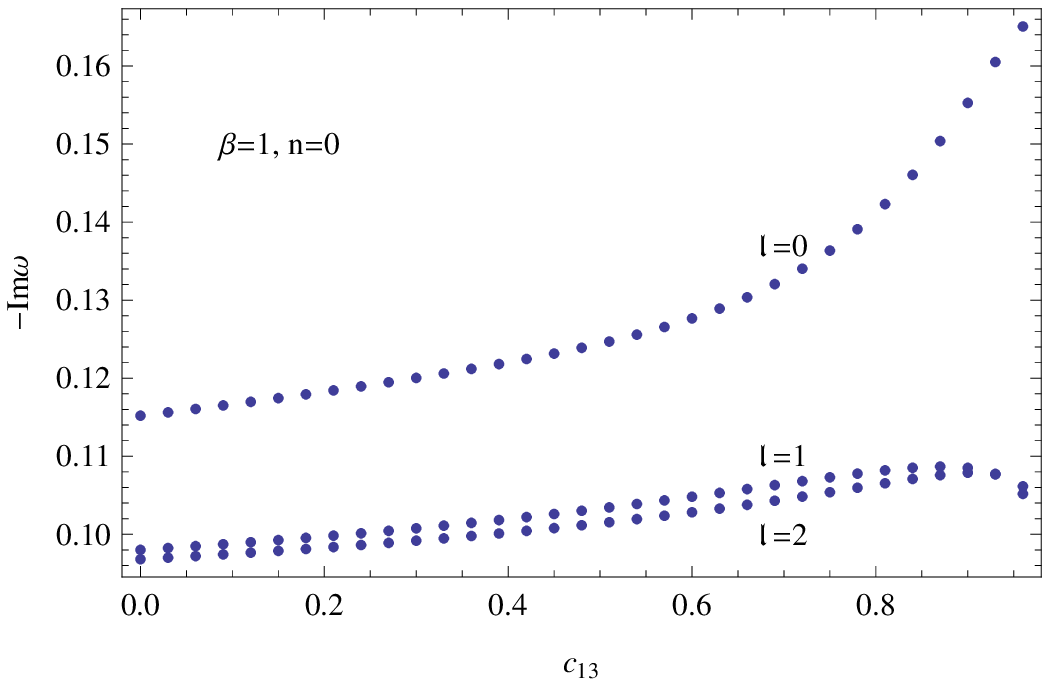}
\caption{The real (left) and imaginary (right) parts of quasinormal frequencies of the scalar field in the background of the first kind aether black hole with different $c_{13}$.}\label{fp22}
 \end{center}
 \end{figure}

\begin{figure}[ht]
\begin{center}
 \includegraphics[width=7cm]{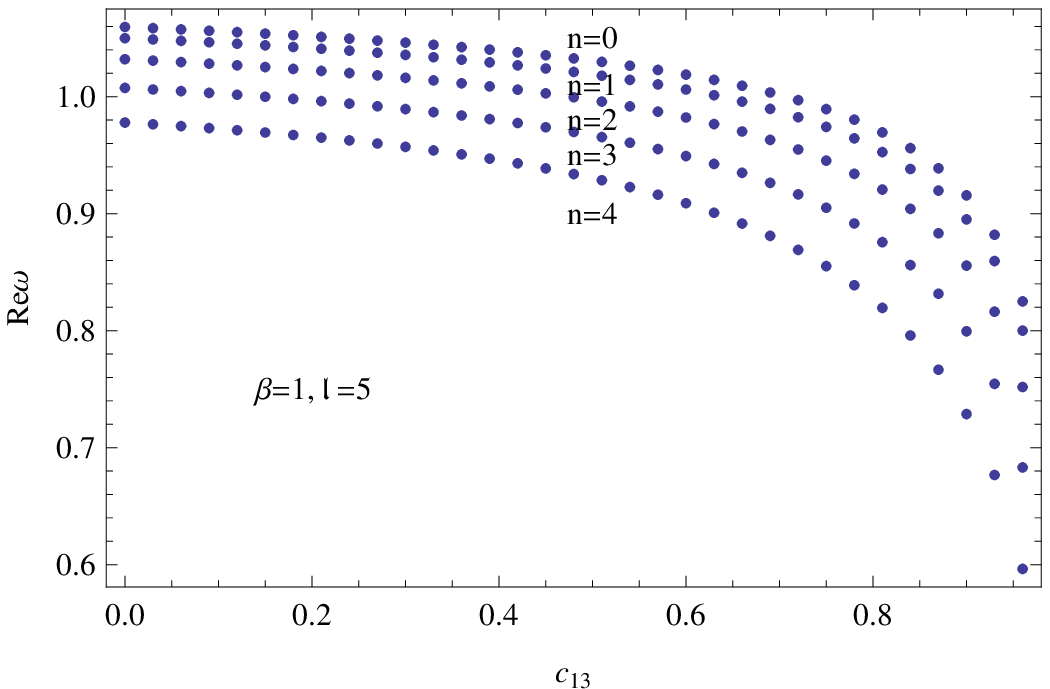}\;\;\;\;\includegraphics[width=7cm]{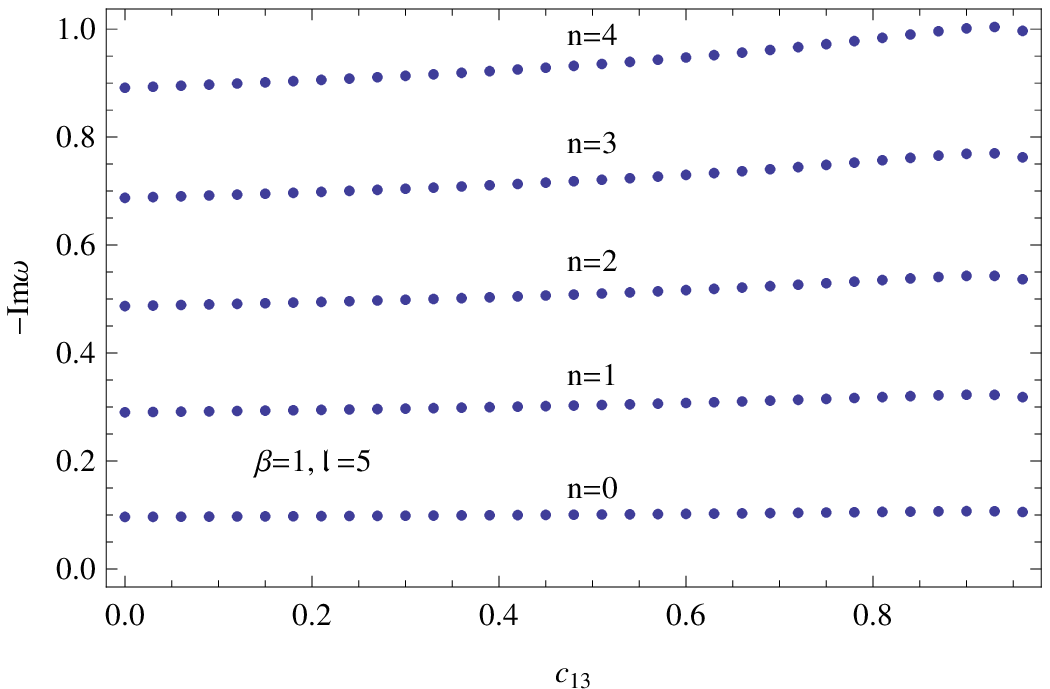}
\caption{The real (left) and imaginary (right) parts of quasinormal frequencies of the scalar field in the background of the first kind aether black hole with different $c_{13}$.}\label{fp23}
 \end{center}
 \end{figure}

\begin{figure}[ht]
\begin{center}
 \includegraphics[width=7cm]{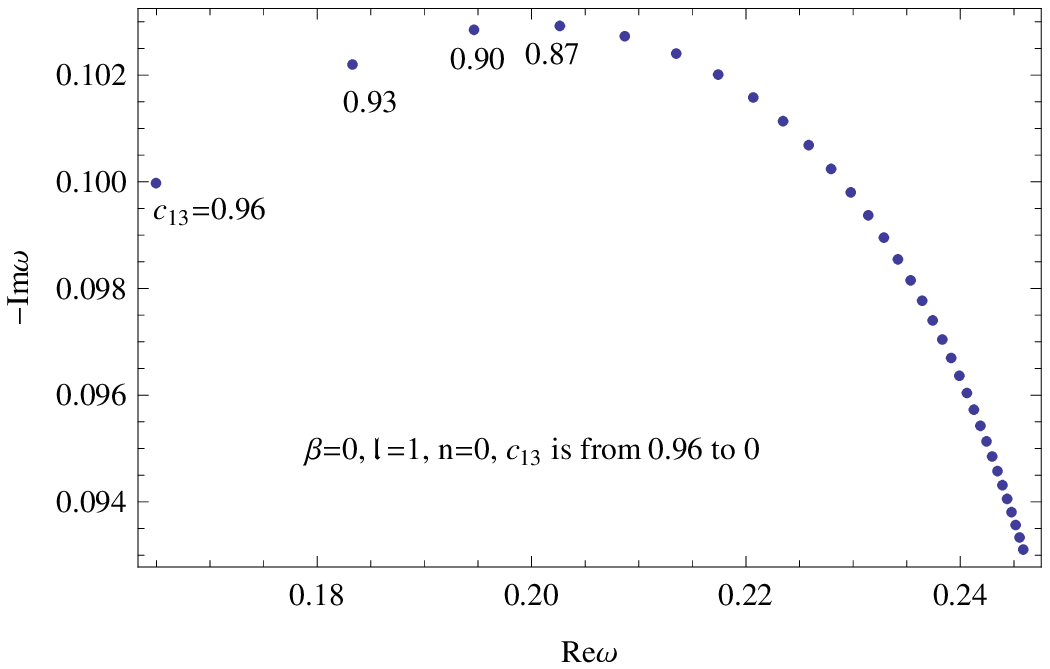}\;\;\;\;\includegraphics[width=7cm]{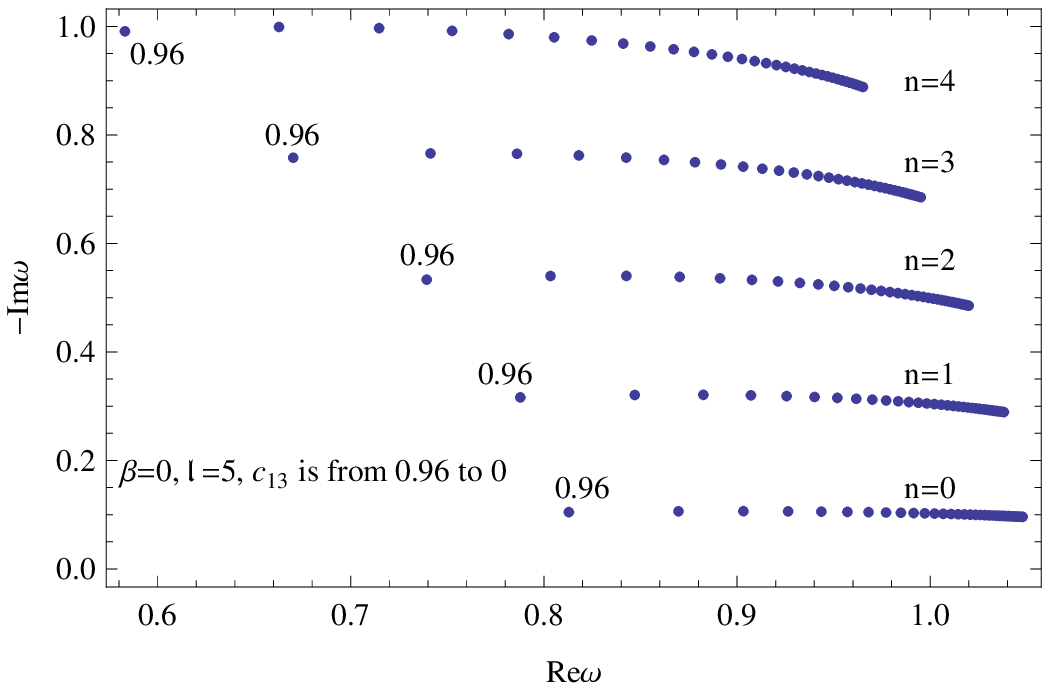}
\caption{The relationship between the real and imaginary parts of quasinormal frequencies of the electromagnetic field in the background of the first kind aether black hole with the decreasing of $c_{13}$.}\label{fp24}
 \end{center}
 \end{figure}

In this section, we study the scalar and electromagnetic field perturbations to the first kind Einstein aether black hole. The scalar  field perturbations are shown in Tab. \ref{tab1} and Fig. \ref{fp2} to \ref{fp23}. The electromagnetic field perturbations are shown in Tab. \ref{tab2} and Fig. \ref{fp24}.

 In Tab. \ref{tab1}, we list the lowest overtone quasinormal modes of massless scalar field for some $l$ with different aether coefficient $c_{13}$. Tab. \ref{tab1} shows that, for fixed $c_{13}$, the real part of frequencies increase and, the absolute imaginary part of them decrease with the angular quantum number $l$. For large $l$, the imaginary parts approach a fixed value.  These properties are similar to the usual black holes and, are also shown in Fig. \ref{fp22}.

Tab \ref{tab1} also shows the derivations from Schwarzschild black hole. For $l=1$, the decrease in Re$\omega$ is about from $0.7$ percent to $17$ percents, while the increase in $-$Im$\omega$ is about from 1 percent to 11 percents, and could be detected by new generation of gravitational antennas. It will help us to seek LV information in nature in low energy scale.

 \begin{table}
 \caption{The lowest overtone $(n=0)$ quasinormal frequencies of the massless scalar field in the first kind aether black hole spacetime.}\label{tab1}
 \begin{center}
\begin{tabular}{cccccc}
\hline \hline $c_{13}$&$\omega(l=0)$&$\omega(l=1)$&$\omega(l=2)$&$\omega(l=3)$&$\omega(l=4)$\\
\hline
 0.00&\;0.104647-0.115197$i$&\;0.291114-0.098001$i$&\;0.483211-0.096805$i$&\; 0.675206-0.096512$i$&\;0.867340-0.096396$i$\\
  0.15&\;0.103976-0.117446$i$&\;0.289524-0.099256$i$&\;0.480578-0.097877$i$&\; 0.671547-0.097541$i$&\;0.862658-0.097409$i$\\
0.30&\;0.101739-0.120032$i$&\;0.287271-0.100767$i$&\; 0.477014-0.099179$i$&\;0.666633-0.098790$i$&\;0.856385-0.098640$i$\\
0.45&\;0.096768-0.123153$i$&\;0.283882-0.102601$i$&\;0.471917-0.100790$i$&\; 0.659659-0.100340$i$&\;0.847507-0.100168$i$\\
0.60&\;0.087386-0.127661$i$&\;0.278310-0.104812$i$&\;0.463995-0.102821$i$&\; 0.648904-0.102303$i$&\;0.833850-0.102107$i$\\
0.75&\;0.072016-0.136350$i$&\;0.267685-0.107300$i$&\;0.449780-0.105389$i$&\; 0.629746-0.104808$i$&\;0.809568-0.104588$i$\\
0.90&\;0.051721-0.155269$i$&\;0.239468-0.108510$i$&\;0.414053-0.107887$i$&\; 0.581804-0.107315$i$&\;0.748823-0.107086$i$\\
\hline\hline
\end{tabular}
\end{center}
\end{table}

 \begin{table}
 \caption{The lowest overtone ($n=0$) quasinormal frequencies of the electromagnetic field in the first kind aether black hole spacetime.}\label{tab2}
 \begin{center}
\begin{tabular}{cccccc}
\hline \hline $c_{13}$ &$\omega(l=1)$&$\omega(l=2)$ &$\omega(l=3)$&$\omega(l=4)$ &$\omega(l=5)$ \\
\hline
 0.00&\;0.245870-0.093106$i$&\;0.457131-0.095065$i$&\;0.656733-0.095631$i$&
 \;0.853018-0.095865$i$&\;1.047870-0.095984$i$\\
 0.15&\;0.243928-0.094312$i$&\;0.454325-0.096122$i$&\;0.652964-0.096654$i$&
 \;0.848255-0.096875$i$&\;1.042100-0.096987$i$\\
 0.30&\;0.241266-0.095728$i$&\;0.450551-0.097388$i$&\;0.647915-0.097889$i$&
 \;0.841883-0.098098$i$&\;1.034390-0.098205$i$\\
  0.45&\;0.237420-0.097401$i$&\;0.445196-0.098932$i$&\;0.640772-0.099409$i$&
  \;0.832880-0.099610$i$&\;1.023490-0.099712$i$\\
   0.60&\;0.231411-0.099372$i$&\;0.436956-0.100837$i$&\;0.629806-0.101314$i$&
   \;0.819066-0.101516$i$&\;1.006780-0.101620$i$\\
 0.75&\;0.220681-0.101581$i$&\;0.422371-0.103156$i$&\;0.610386-0.103695$i$&
 \;0.794587-0.103926$i$&\;0.977154-0.104044$i$\\
  0.90&\;0.194630-0.102849$i$&\;0.386513-0.105074$i$&\;0.562260-0.105885$i$&
  \;0.733694-0.106235$i$&\;0.903309-0.106415$i$\\
\hline\hline
\end{tabular}
\end{center}
\end{table}

In another side, for the fixed angular number $l$, or overtone number $n$ with different $c_{13}$ (small $c_{13}$), Tab. \ref{tab1}, Fig. \ref{fp2} and Fig. \ref{fp23} show that the real part of frequencies decrease, and the absolute imaginary ones increase with $c_{13}$ firstly to $c_{13}=0.87$ and then decrease on the contrary, which is different from that of the non-reduced aether black hole \cite{konoplya}, where both all increase with $c_1$.

For different overtone numbers $n$, Fig. \ref{fp2} and Fig. \ref{fp23} show that the real parts decrease and the absolute imaginary ones increase with $n$, which is the same as that of Schwarzschild black hole.

For angular number $l=0$ and $l>0$, Fig. \ref{fp22} shows an unusual behavior. When $l>0$, the absolute imaginary part of frequencies increases for small $c_{13}$, and then decrease in the region of large $c_{13}$. However when $l=0$, it increases for all $c_{13}$. This behavior is similar to the case of deformed Ho\v{r}ava-Lifshitz black hole where the real part one increases for $l>0$ and only decreases for $l=0$ with the coefficient $\alpha$ \cite{chen2010}. It is related to its unusual potential behavior Fig. \ref{fp1}.

By compared to Reissner-Norstr\"{o}m black hole, Fig. \ref{fp2} and \ref{fp22} show us a similar behavior that the absolute imaginary part of frequencies increases for small parameter $c_{13}$ or $Q$, and then decrease in the region of large parameter \cite{konoplya}. The only difference is that the real part decreases here for all $c_{13}$ and increases there for all $Q$.

 For fixed $l$, Tab. \ref{tab2} and Fig. \ref{fp24} shows us that the behavior of electromagnetic perturbation frequencies is similar to that of the scalar case. The real part of frequencies decreases for all $c_{13}$ and, the absolute imaginary one increases for small $c_{13}$, and then decrease in the region of large $c_{13}$. For fixed $c_{13}$, Tab. \ref{tab2} shows that both the real and the absolute imaginary parts increase with $l$.
And the real and absolute imaginary parts of electromagnetic field are smaller than those corresponding value of scalar field.

By compared to another LV model --- the QED-extension limit of standard model extension (SME, see Appendix for more detail), the above scalar and electromagnetic field QNMs properties with $c_{13}$ are similar to Dirac field QNMs with LV coefficient $b$ \cite{chen2006}, i.e., the real part decreases while the absolute imaginary part increases with the given LV coefficient. In the theory of QED-extension limit of SME, local LV coefficient $b_\mu$ is introduced in a matter sector, while in Einstein aether theory, local LV is in a gravity sector. For the former, LV matter perturbs to LI black hole --- Schwarzschild black hole and produces QNMs. For the latter, LI matter perturbs to LV black hole --- Einstein aether black hole and then produces QNMs. These similarities between different backgrounds may imply some common property of LV coefficient on QNMs, i.e., in presence of LV, the perturbation field oscillation damps more rapidly, and its period becomes longer.

\section{Quasinormal modes for the second kind aether black hole}

\begin{figure}[ht]
\begin{center}
\includegraphics[width=7.0cm]{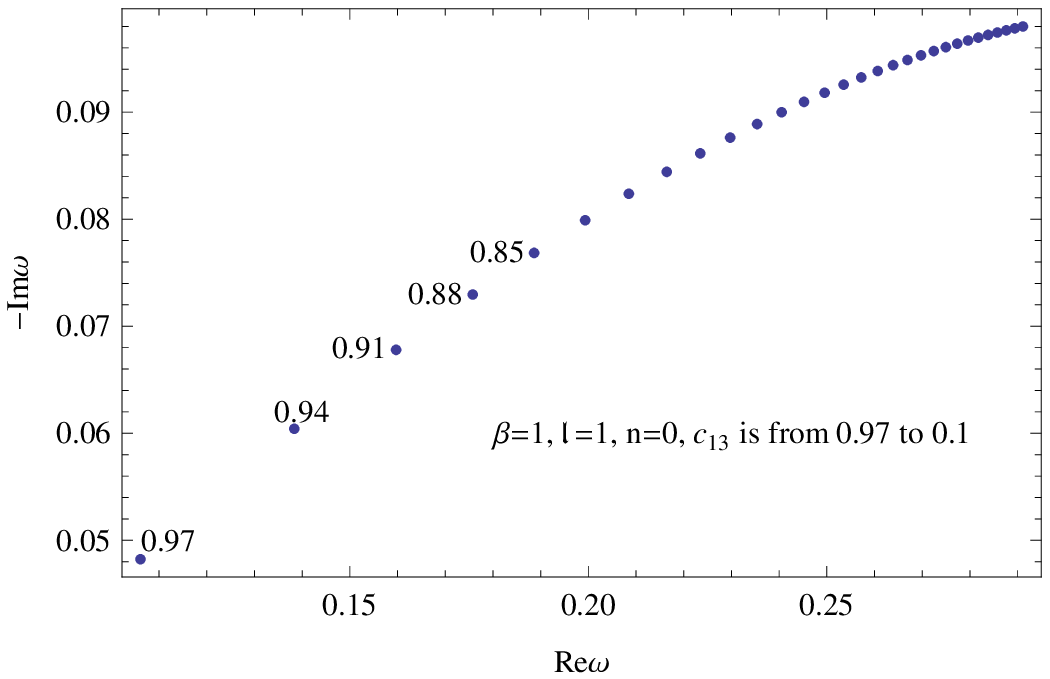}\;\;\;\;
 \includegraphics[width=7.0cm]{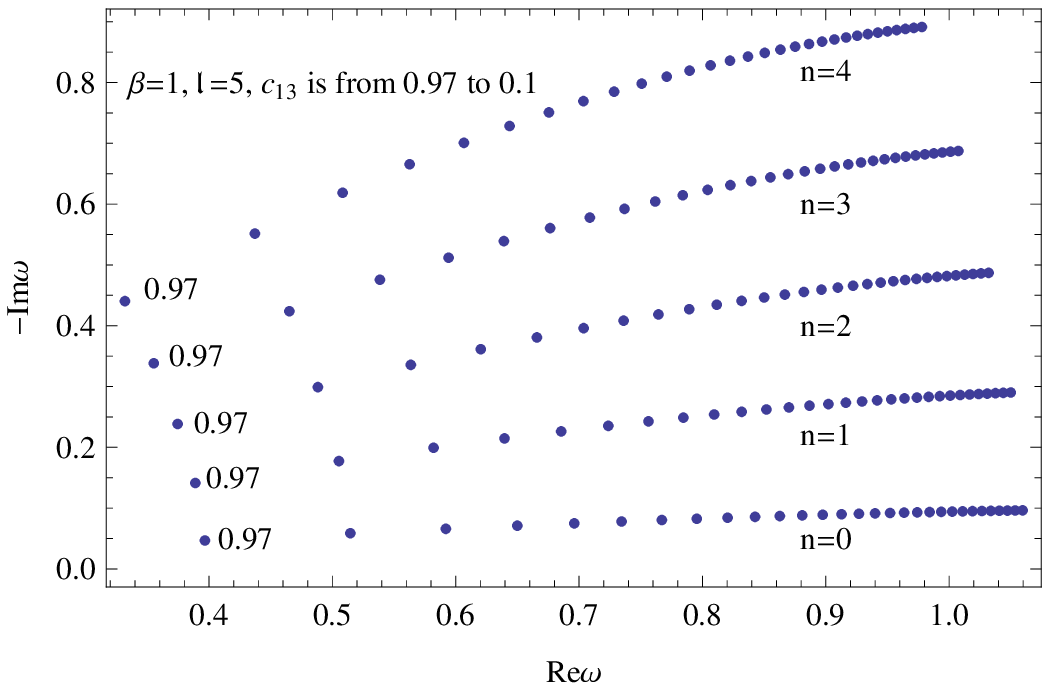}
\caption{The relationship between the real and imaginary parts of quasinormal frequencies of the scalar field in the background of the second kind aether black hole with the decreasing of $c_{13}$. }\label{fp3}
 \end{center}
 \end{figure}

\begin{figure}[ht]
\begin{center}
\includegraphics[width=7.0cm]{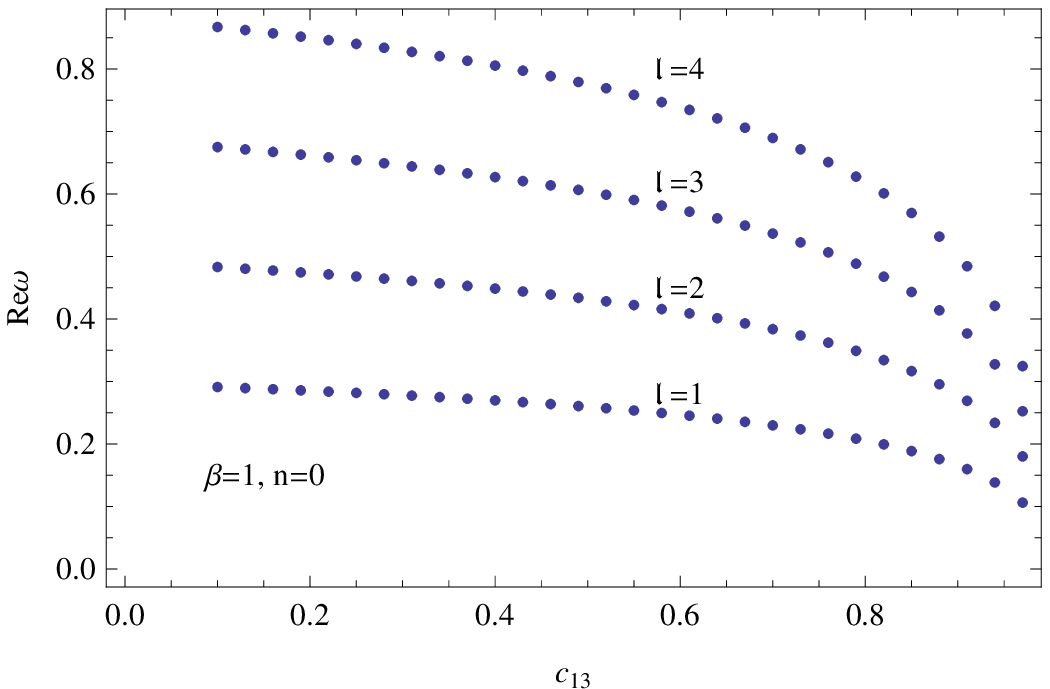}\;\;\;\;
 \includegraphics[width=7.0cm]{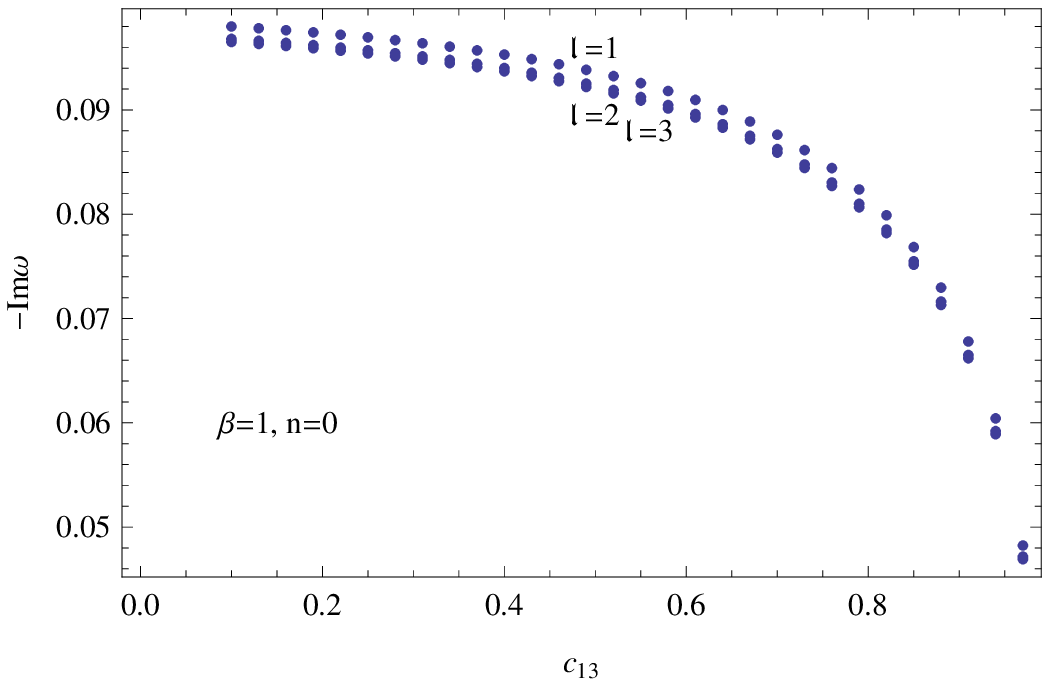}
\caption{The real (left) and imaginary (right) parts of quasinormal frequencies of the scalar field in the background of the second kind aether black hole with different $c_{13}$. }\label{fp32}
 \end{center}
 \end{figure}

\begin{figure}[ht]
\begin{center}
\includegraphics[width=7.0cm]{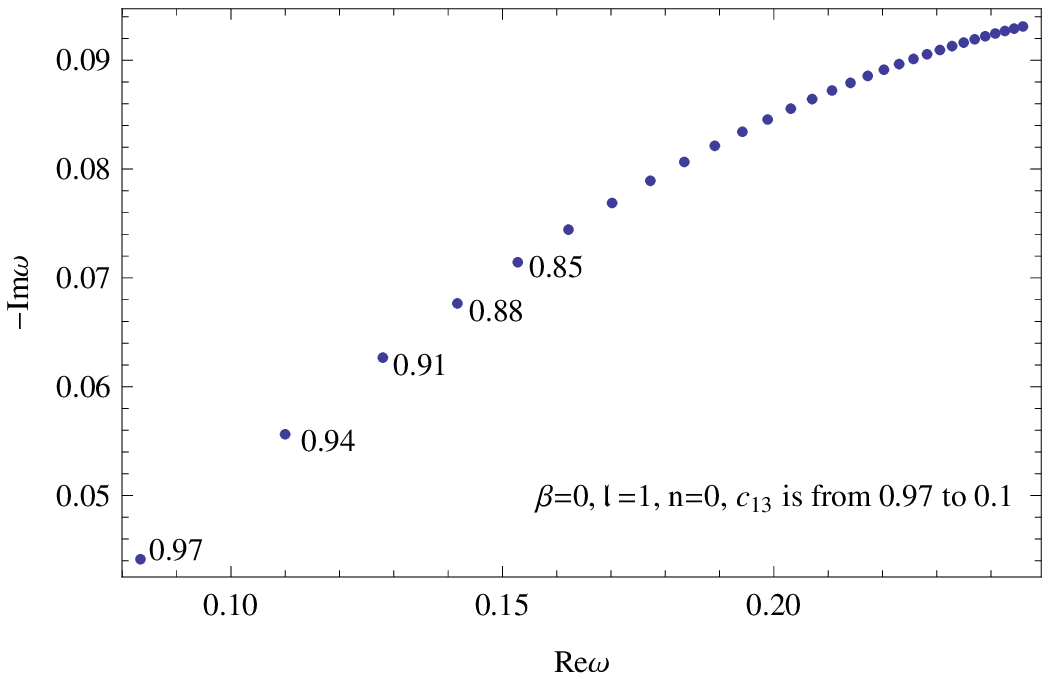}\;\;\;\;\;
 \includegraphics[width=7.0cm]{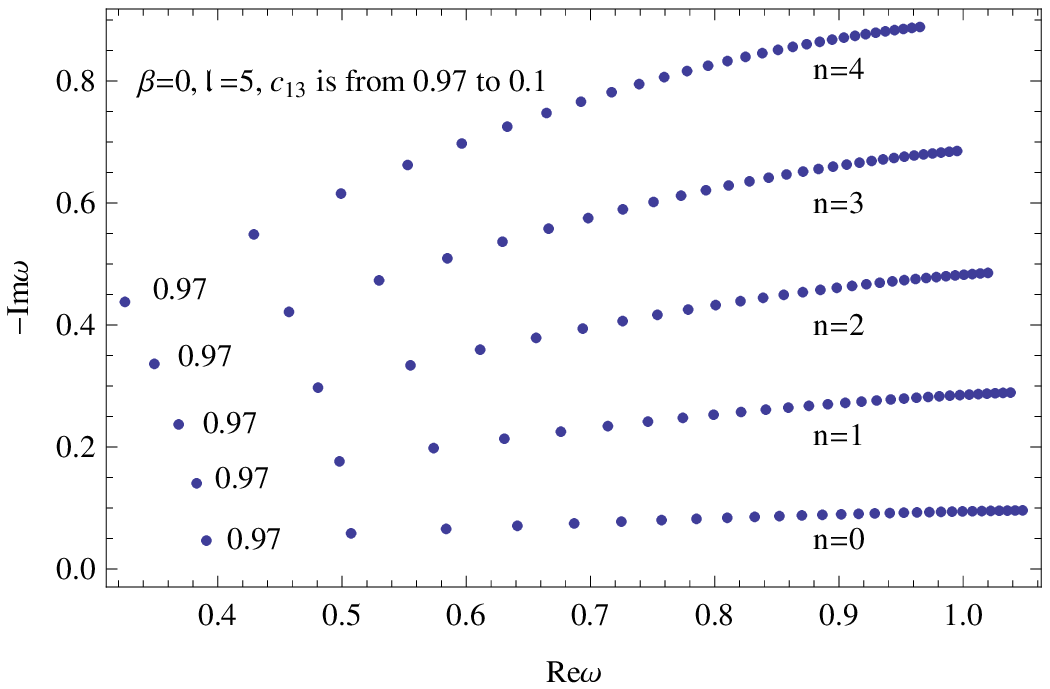}
\caption{The relationship between the real and imaginary parts of quasinormal frequencies of the electromagnetic field in the background of the second kind aether black hole with the decreasing of $c_{13}$ } \label{fp33}
 \end{center}
 \end{figure}

In this section, we study the scalar and electromagnetic field perturbations to the second kind of Einstein aether black hole with fixed $c_{14}=0.2$.  The scalar field perturbations are shown in Tab. \ref{tab3}, Fig. \ref{fp3} and Fig. \ref{fp32}. The electromagnetic field perturbations are shown in \ref{tab4} and Fig. \ref{fp33}.

 In Tab. \ref{tab3}, we list the lowest overtone quasinormal modes of massless scalar field for some $l$ with different aether coefficient $c_{13}$. Tab. \ref{tab3} shows that, for fixed $c_{13}$, the real part of frequencies increase and, the absolute imaginary part of them decrease with the angular quantum number $l$. For large $l$, the imaginary part approach a fixed value which is similar to the usual black holes and,  also shown in Fig. \ref{fp32}.

Tab \ref{tab3} also shows the derivations from Schwarzschild black hole. For $l=1$, the decrease in Re$\omega$ is about from $3$ percents to $35$ percents, while the decrease in $-$Im$\omega$ is about from 1 percent to 21 percents, both are bigger than the first kind aether black hole, and could be detected by new generation of gravitational antennas.

 In another side, for the fixed angular number $l$, or overtone number $n$,  Tab. \ref{tab3}, Fig. \ref{fp3} and Fig. \ref{fp33} show that both the real and  the absolute imaginary parts of frequencies all decrease with $c_{13}$ increasing, which is completely different from that of the non-reduced aether black hole \cite{konoplya}, where both all increase with $c_1$. This property of both decrease is similar to that of Einstein-Born-Infeld black hole \cite{fernando,chen2010}.

 \begin{table}
 \caption{The lowest overtone $(n=0)$ quasinormal frequencies of the massless scalar field in the second kind aether black hole spacetime with fixed $c_{14}=0.2$.}\label{tab3}
 \begin{center}
\begin{tabular}{cccccc}
\hline \hline $c_{13}$ &$\omega(l=0)$&$\omega(l=1)$&$\omega(l=2)$&$\omega(l=3)$&$\omega(l=4)$\\
\hline
 0.10&\;0.104647-0.115197$i$&\;0.291114-0.098001$i$&\;0.483211-0.096805$i$&\; 0.675206-0.096512$i$&\;0.867340-0.096396$i$\\
  0.25&\;0.100755-0.114893$i$&\;0.281760-0.096962$i$&\;0.468061-0.095718$i$&\; 0.654122-0.095413$i$&\;0.840293-0.095293$i$\\
0.40&\;0.095828-0.114071$i$&\;0.269748-0.095309$i$&\;0.448616-0.094016$i$&\;
0.627061-0.093699$i$&\;0.805577-0.093575$i$\\
0.55&\;0.089374-0.112234$i$&\;0.253524-0.092567$i$&\;0.422340-0.091228$i$&\;
0.590489-0.090899$i$&\;0.758658-0.090770$i$\\
0.70&\;0.080354-0.108123$i$&\;0.229737-0.087618$i$&\;0.383738-0.086246$i$&\; 0.536739-0.085908$i$&\;0.689690-0.085775$i$\\
0.85&\;0.065688-0.097327$i$&\;0.188631-0.076846$i$&\;0.316653-0.075488$i$&\; 0.443244-0.075155$i$&\;0.569686-0.075023$i$\\
\hline\hline
\end{tabular}
\end{center}
\end{table}

For different overtone numbers $n$, Fig. \ref{fp3} and Fig. \ref{fp33} show that the real parts decrease and the absolute imaginary ones increase with $n$, which is the same as that of Schwarzschild black hole.

 \begin{table}
 \caption{The lowest overtone ($n=0$)  quasinormal frequencies of the electromagnetic field in the second kind aether black hole spacetime with fixed $c_{14}=0.2$.}\label{tab4}
 \begin{center}
\begin{tabular}{cccccc}
\hline \hline $c_{13}$&$\omega(l=1)$&$\omega(l=2)$&$\omega(l=3)$&$\omega(l=4)$&$\omega(l=5)$\\
\hline
 0.10&\;0.245870-0.093106$i$&\;0.457131-0.095065$i$&\;0.656733-0.095631$i$&\;
 0.853018-0.095865$i$&\;1.047870-0.095984$i$\\
 0.25&\;0.236985-0.091929$i$&\;0.442267-0.093924$i$&\;0.635855-0.094505$i$&\;
  0.826131-0.094746$i$&\;1.014980-0.094868$i$\\
 0.40&\;0.225711-0.090122$i$&\;0.423262-0.092160$i$&\;0.609107-0.092759$i$&\;
 0.791659-0.093008$i$&\;0.972796-0.093135$i$\\
  0.55&\;0.210705-0.087219$i$&\;0.397700-0.089304$i$&\;0.573042-0.089924$i$&\;
  0.745134-0.090183$i$&\;0.915833-0.090314$i$\\
   0.70&\;0.189117-0.082138$i$&\;0.360372-0.084257$i$&\;0.520194-0.084899$i$&\;
   0.676866-0.085167$i$&\;0.832193-0.085304$i$\\
 0.85&\;0.152828-0.071437$i$&\;0.296066-0.073499$i$&\;0.428661-0.074143$i$&\;
 0.558382-0.074414$i$&\;0.686885-0.074552$i$\\
\hline\hline
\end{tabular}
\end{center}
\end{table}

For fixed $l$, Tab. \ref{tab4} and Fig. \ref{fp33} shows us that the behavior of electromagnetic perturbation frequencies is similar to that of the scalar case, i.e., both the real part and the absolute imaginary one of frequencies decreases for all $c_{13}$. The only difference is that the real and absolute imaginary parts of electromagnetic field are smaller than those corresponding value of scalar field. For fixed $c_{13}$, both the real and the imaginary parts increase for all $l$.

\section{Summary}

The gravitational consequence of local Lorentz violation should show itself in radiative processes around black holes. The significant difference between Einstein and Einstein aether theories can show itself in derivation of the characteristic QNMs of black hole mergers from their Schwarzschild case.

In this paper, we study on QNMs of the scalar and electromagnetic field perturbations to Einstein aether black holes. There exist a series of single parameter $c_{13}$ black holes solutions: the first and the second kind aether black hole, instead of four coefficients there in Einstein aether theory.

For the effective potential, when $c_{13}$ increases, its turning point always becomes larger, and the value of its peak becomes lower except a special case that for the first aether black hole with angular quantum number $l=0$, where its peak becomes higher.

For the three kinds, the first, the second and the non-reduced aether black holes, their QNMs are different from each other that show their complexity. For the non-reduced aether hole \cite{konoplya}, both real part and the absolute value of imaginary part of QNMs (both scalar and electromagnetic fields) increase with $c_1$. On the contrary for the second kind aether black hole, both decrease with $c_{13}$, that are similar to the scalar field QNMs of Einstein-Born-Infeld black holes with the Born-Infeld parameter $b$ \cite{chen2010}.

For the first kind aether black hole, the real part of scalar and electromagnetic QNMs becomes smaller with all $c_{13}$ increase. The absolute value of imaginary part of QNMs becomes bigger with small $c_{13}$ increase, and then decreases with big $c_{13}$.  These properties with $c_{13}$ are similar to the behaviors of Dirac QNMs with LV coefficient $b$ \cite{chen2006}. These similarities between different backgrounds may imply some connections between Einstein aether theory and the QED-extension limit of SME, i.e., LV in gravity sector and LV in matter sector will make quasinormal ringing of black holes damping more rapidly and its period becoming longer.

Compared to Schwarzschild black hole, both the first and the second kind aether black holes have larger damping rate and smaller real oscillation frequency of QNMs. And the differences are from 0.7 percent to 35 percents, those could be detected by new generation of gravitational antennas. If the breaking of Lorentz symmetry is not very small, the derivation of QNMs from Schwarzschild values might be observed in the near future gravitational wave events.

\begin{acknowledgments}
This work was supported by the National Natural Science Foundation
of China (grant No. 11247013), Hunan Provincial Natural Science Foundation of China (grant No. 2015JJ2085), and the fund under grant No. QSQC1203.
\end{acknowledgments}

\appendix\section{Standard Model Extension}

Invariance under Lorentz transformation is one of the pillars of both Einstein's General Relativity (GR) and the Standard Model (SM) of particle physics. GR describes gravitation at the classical level, while SM encompasses all other phenomena involving the basic particles and forces down to the quantum level \cite{colladay}. They are expected to merge at the Plank scale into a single unified and quantum-consistent description of nature. The underlying unified quantum gravity theory indicates the Lorentz symmetry is only an approximate symmetry emerging at low energies and will be violated at ultrahigh energies. Any observable signals of LV can be described using effectively field theory. An effective field theory construction known as the gravitational standard model extension (SME) provides such a comprehensive framework that contains known physics along with all possible LV effects. So that SME isn't a specific model, but a construction ideally suited for a broad search and having the power to predict the outcome of relevant experiments \cite{tasson}.

The background of SME is Riemann-Cartan geometry which allows for nonzero vacuum quantities that violate local Lorentz invariance but preserve general coordinate  invariance. The usual Riemann spacetime of GR can be recovered in the zero torsion limit.
SME can be constructed from the action of GR and SM by adding all local LV and coordinated-independent  terms, perhaps together with suppressed higher-order terms, and they incorporate general CPT (Charge, Parity and Time symmetry) violation \cite{kostelecky}.
Each SME term comes with a coefficient for LV governing the size of the associated experimental signals.
The action $S_{\text{SME}}$ for the full SME can be expressed as a sum of partial actions. These actions can be split into a matter sector and a gravity sector,
\begin{eqnarray}
 S_{\text{SME}}=S_{\text{matter}}+S_{\text{gravity}}+\cdots.
\end{eqnarray}
In curved spacetime, the matter sector naturally couples to gravitation.
The matter sector is given by
\begin{eqnarray}\label{sme}
 S_{\text{matter}}=S_{\text{SM}}+S_{\psi}+S_{A}.
\end{eqnarray}
The term $S_{\text{SM}}$ is the SM action, modified by the addition of gravitational couplings and containing all Lorentz- and CPT-violating terms that involve SM fields and dominate at low energies. SM fields include the three charged leptons, the three neutrinos, the six quark flavors, Higgs, gauge and Yukawa couplings. The terms $S_{\psi}$ and $S_{\text{A}}$ are the QED-extension (Quantum Electrodynamics) limit of SME, Dirac fermion $\psi$ and the photon $A_\mu$, respectively. The term $S_{\text{gravity}}$ represents the pure-gravity sector incorporating possible Lorentz and CPT violation. It is easy to see that Einstein aether theory is a specifical model of $S_{\text{gravity}}$. The ellipsis represents contributions to $S_{\text{SME}}$ that are of higher order at low energies.

The fermion sector $S_{\psi}$
in (\ref{sme}) can be explicitly expressed as \cite{chen2006,colladay}
\begin{eqnarray}
S_{\psi}=\int d^4x\sqrt{-g}(\frac{1}{2}i e^{\mu}_a \overline{\psi}
\Gamma^a \overleftrightarrow{D_{\mu}}\psi-\overline{\psi} M^*
\psi),\label{act}
\end{eqnarray}
where $e^{\mu}_{\;\;a}$ is the inverse of the vierbein
$e^a_{\;\;\mu}$. The symbols $\Gamma^a$ and $M^*$ are
\begin{eqnarray}
\Gamma^a\equiv \gamma^a-c_{\mu\nu}e^{\nu a}e^{\mu}_{\;\;
b}\gamma^b-d_{\mu\nu}e^{\nu a}e^{\mu}_{\;\;
b}\gamma_5\gamma^b-e_{\mu}e^{\mu a}-i f_{\mu}e^{\mu
a}\gamma_5-\frac{1}{2}g_{\lambda\mu\nu}e^{\nu a}e^{\lambda}_{\;\;
b}e^{\mu}_{\;\; c}\sigma^{bc}\label{gam1},
\end{eqnarray}
and
\begin{eqnarray}
M^*\equiv m +i m_5\gamma_5+a_{\mu}e^{\mu}_{\;\;a}\gamma^a
+b_{\mu}e^{\mu}_{\;\;a}\gamma_5\gamma^a+\frac{1}{2}H_{\mu\nu}e^{\mu}_{\;\;a}e^{\nu}_{\;\;b}\sigma^{ab}.\label{gam2}
\end{eqnarray}
The first terms of Eqs.(\ref{gam1}) and (\ref{gam2}) lead to the
usual Lorentz invariant kinetic term and mass for the Dirac field.
The parameters $a_{\mu}$, $b_{\mu}$, $c_{\mu\nu}$, $d_{\mu\nu}$,
$e_{\mu}$, $f_{\mu}$, $ g_{\lambda\mu\nu}$, $H_{\mu\nu}$ are
Lorentz violating coefficients which arise from nonzero vacuum
expectation values of tensor quantities and comprehensive describe
effects of Lorentz violation on  the behavior of particles coupling
to these tensor fields.

From the action (\ref{act}), we can obtain that due to
presence of Lorentz violating coefficients, Dirac equation must be
modified. According to the variation of $\overline{\psi}$ in the action
(\ref{act}), we find that the massless Dirac equation only
containing the CPT and Lorentz covariance breaking kinetic term
associated with an axial-vector $b_{\mu}$ field in the curve
spacetime can be expressed as
\begin{eqnarray}
[i\gamma^ae_{a}^{\;\;\mu}(\partial_{\mu}+\Gamma_{\mu})-b_{\mu}e_{a}^{\;\;\mu}\gamma_{5}\gamma^a]\Psi=0,\label{deq1}
\end{eqnarray}
where
\begin{eqnarray}
\gamma^{0}=\left(
\begin{array}{c}I\;\;\;\;0\\
0 \;\; -I
\end{array}\right),\;\;\;\;\;
\gamma^{i}=\left(
\begin{array}{c}0\;\;\;\;\sigma^{i}\\
-\sigma^{i} \;\;0
\end{array}\right),\;\;\;\;\;
\gamma_{5}=i\gamma^{0}\gamma^{1}\gamma^{2}\gamma^{3}=\left(
\begin{array}{c}0\;\;\;\;I\\
I \;\;\;\; 0
\end{array}\right).
\end{eqnarray}
It is reasonable for us
to assume that the axial-vector $b_{\mu}$ field does not change the
background metric since the Lorentz violation is very small. For convenience, we take $b_{\mu}$ as a non-zero
timelike vector $(\frac{b}{r^2},0,0,0)$, where $b$ is a constant. The vierbein of Schwarzschild spacetime  can be defined as
\begin{eqnarray}
e^a_{\;\;\mu}=(\sqrt{1-2M/r},
\frac{1}{\sqrt{1-2M/r}}, \;\; r, \;\;
r\sin{\theta})\label{vier1}.
\end{eqnarray}
And then by using WKB method, Chen {\it et al} \cite{chen2006} obtained the influence of LV on Dirac QNMs in Schwarzschild spacetime. They found that at fundamental overtone, the real part decreases linearly as the parameter $b$ increases, while for the larger multiple moment $k$,
the absolute imaginary part increases with $b$, which means that presence of Lorentz violation makes Dirac field damps more rapidly. These behaviors with LV coefficient are similar to those of the first kind Einstein aether black hole.

\section{Accuracy of Wkb method}

The Schr\"{o}dinger-like wave equation (\ref{schrodinger}) with the effective potential
(\ref{potential}) is not solvable analytically. A numerical integration of it requires selecting a value for the complex frequency, then integrating and checking whether the boundary conditions for QNMs are satisfied. Since those conditions are not met in general, the complex frequency plane must be surveyed for discrete values. So this technique is time consuming and costly. However for a semianalytic method, there is an accuracy problem which will be discussed in this section.

As is well known, the accuracy of the WKB approximation should increase with multipole momentum $l$, then only the low lying modes of the lowest overtone are considered here. Firstly the lowest three modes of the scalar and electromagnetic field perturbations of Schwarzschild black hole are shown in Tab. \ref{tab11} via three methods: the Schutz-Will approximation (the first order WKB) \cite{schutz}, the third order WKB \cite{iyer1,iyer2} and the numerical technique \cite{leaver}. The formula for the complex quasinormal frequencies in the Schutz-Will approximation is given by
 \begin{eqnarray}
\omega^2=V_0-i\Big(n+\frac{1}{2}\Big)\sqrt{-2V_0''},
\end{eqnarray}
with
 \begin{eqnarray}
V_0^{(m)}=\frac{d^mV_i}{dr_*^m}\Big|_{r_*(r_p)},
\end{eqnarray}
where $n$ is overtone number and $r_p$ is the turning point value of polar coordinate $r$ at which  the effective potential reaches its maximum (\ref{potential}).

From Tab. \ref{tab11}, one can see that the results via the 3rd WKB method have higher accuracy than the 1st one.
Then to the correctness of the present data in the maintext, the simplest way is to compare them to those obtained by the 1st WKB method. So the three lowest modes for both types of the aether black hole and for the scalar and electromagnetic perturbations via this method are listed in Tab. \ref{tab12} and \ref{tab13}.

To the first kind aether black hole, by comparing Tab. \ref{tab12} to Tabs. \ref{tab1} and \ref{tab2}, one can see that the behavior of these data is the same as those, i.e., for fixed $c_{13}$, the real part of frequencies increase and, the absolute imaginary parts decrease for the scalar field, increase for the electromagnetic one with the angular quantum number $l$; for the fixed angular number $l$ (except $l=0$ case), the real part of frequencies decrease, and the absolute imaginary ones increase with $c_{13}$. To the case of fixed $l=0$, the behavior of the real part of scalar perturbation modes is different from Tab. \ref{tab1}, which may be due to its unusual potential.

To the second kind aether black hole, by comparing Tab. \ref{tab13} to Tabs. \ref{tab3} and \ref{tab4}, one can see that the behavior of these data is the same as those, i.e., for fixed $c_{13}$, the real part of frequencies increase and, the absolute imaginary part of them decrease for the scalar field, increase for the electromagnetic field with $l$; for the fixed angular number $l$, both the real part and the absolute imaginary of frequencies decrease with $c_{13}$. Therefore, both methods give approximately the same modes $\omega$.

\begin{table}
 \caption{The three lowest modes $(n=0)$ quasinormal frequencies of the massless scalar (above) and electromagnetic (lower) field in the Schwarzschild black hole spacetime. The numerical results are from \cite{leaver}}\label{tab11}
 \begin{center}
\begin{tabular}{cccc}
\hline \hline $l$ & 1st WKB & 3rd WKB & numerical \\
\hline
 0&\;0.189785-0.098240$i$&\;0.104647-0.115197$i$&\;0.1105-0.1049$i$\\
  1&\;0.329434-0.096256$i$&\;0.291114-0.098001$i$&\;0.2929-0.0977$i$\\
 2&\;0.506317-0.096123$i$&\;0.483211-0.096805$i$&\;0.4836-0.0968$i$\\
\hline
 1&\;0.287050-0.091235$i$&\;0.245870-0.093106$i$&\;0.2483-0.0925$i$\\ 2&\;0.480754-0.094354$i$&\;0.457131-0.095065$i$&\;0.4576-0.0950$i$\\
 3&\;0.673438-0.095258$i$&\;0.656733-0.095631$i$&\;0.6569-0.0956$i$\\
\hline\hline
\end{tabular}
\end{center}
\end{table}
\begin{table}
 \caption{The three lowest modes $(n=0)$ quasinormal frequencies of the massless scalar (above) and electromagnetic (lower) field in the first kind aether black hole spacetime via the 1st WKB method.}\label{tab12}
 \begin{center}
\begin{tabular}{cccccccc}
\hline \hline $l$&$\omega(c_{13}=0)$ & $\omega(c_{13}=0.15)$ & $\omega(c_{13}=0.3)$ & $\omega(c_{13}=0.45)$
&$\omega(c_{13}=0.6)$&$\omega(c_{13}=0.75)$&$\omega(c_{13}=0.9)$ \\
\hline
 0&\;0.1898-0.0982$i$&\;0.1910-0.0995$i$&\;0.1925-0.1012$i$&0.1945-0.1036$i$&0.1971-0.1070$i$
 &0.2007-0.1119$i$&0.2035-0.1188$i$\\
  1&\;0.3294-0.0963$i$&\;0.3286-0.0972$i$&\;0.3274-0.0983$i$&0.3257-0.0998$i$&0.3229-0.1019$i$
  &0.3176-0.1047$i$&0.3025-0.1085$i$\\
 2&\;0.5063-0.0961$i$&\;0.5042-0.0971$i$&\;0.5013-0.0982$i$&0.4971-0.0997$i$&0.4907-0.1015$i$
 &0.4789-0.1040$i$&0.4485-0.1067$i$\\
\hline
 1&\;0.2871-0.0912$i$&\;0.2859-0.0920$i$&\;0.2844-0.0930$i$&0.2822-0.0942$i$&0.2787-0.0956$i$
 &0.2723-0.0974$i$&0.2554-0.0985$i$\\ 2&\;0.4808-0.0944$i$&\;0.4785-0.0953$i$&\;0.4754-0.0964$i$&0.4710-0.0977$i$&0.4642-0.0995$i$
 &0.4521-0.1016$i$&0.4211-0.1035$i$\\
 3&\;0.6734-0.0953$i$&\;0.6700-0.0962$i$&\;0.6655-0.0974$i$&0.6591-0.0988$i$&0.6491-0.1006$i$
 &0.6314-0.1029$i$&0.5867-0.1051$i$\\
\hline\hline
\end{tabular}
\end{center}
\end{table}
\begin{table}
 \caption{The three lowest modes $(n=0)$ quasinormal frequencies of the massless scalar (above) and electromagnetic (lower) field in the second kind aether black hole spacetime via the 1st WKB method.}\label{tab13}
 \begin{center}
\begin{tabular}{ccccccc}
\hline \hline $l$&$\omega(c_{13}=0.1)$ & $\omega(c_{13}=0.25)$ & $\omega(c_{13}=0.4)$ & $\omega(c_{13}=0.55)$
&$\omega(c_{13}=0.7)$&$\omega(c_{13}=0.85)$\\
\hline
 0&\;0.1898-0.0982$i$&\;0.1869-0.0975$i$&\;0.1827-0.0963$i$&0.1764-0.0940$i$&0.1657-0.0895$i$
 &0.1438-0.0792$i$\\
  1&\;0.3294-0.0963$i$&\;0.3206-0.0952$i$&\;0.3091-0.0936$i$&0.2932-0.0909$i$&0.2692-0.0861$i$
  &0.2260-0.0756$i$\\
 2&\;0.5063-0.0961$i$&\;0.4914-0.0950$i$&\;0.4722-0.0933$i$&0.4460-0.0905$i$&0.4072-0.0856$i$
 &0.3386-0.0749$i$\\
\hline
 1&\;0.2871-0.0912$i$&\;0.2786-0.0900$i$&\;0.2678-0.0882$i$&0.2529-0.0853$i$&0.2309-0.0802$i$
 &0.1920-0.0697$i$\\ 2&\;0.4808-0.0944$i$&\;0.4661-0.0932$i$&\;0.4473-0.0914$i$&0.4218-0.0886$i$&0.3842-0.0835$i$
 &0.3183-0.0728$i$\\
 3&\;0.6734-0.0953$i$&\;0.6527-0.0941$i$&\;0.6261-0.0924$i$&0.5901-0.0895$i$&0.5370-0.0845$i$
 &0.4443-0.0738$i$\\
\hline\hline
\end{tabular}
\end{center}
\end{table}

\end{document}